\documentclass{article}

\usepackage[dvips]{graphicx}

\usepackage{amssymb,amsfonts,amsmath}

\begin{document}

\title{Influence of network dynamics on the spread of sexually transmitted diseases}

\author{Sebasti\'an Risau-Gusman \\ Consejo Nacional de Investigaciones Cient\'{\i}ficas y T\'ecnicas, \\
Centro At\'omico Bariloche, 8400 San Carlos de Bariloche, Argentina}

\date{} 

\maketitle

\begin{abstract} Network epidemiology often assumes that the relationships defining the social network of a population are static. The dynamics of relationships is only taken indirectly into account, by assuming that the relevant information to study epidemic spread is encoded in the network obtained by considering numbers of partners accumulated over periods of time roughly proportional to the infectious period of the disease at hand. On the other hand, models explicitly including social dynamics are often too schematic to provide a reasonable representation of a real population, or so detailed that no general conclusions can be drawn from them. Here we present a model of social dynamics that is general enough that its parameters can be obtained by fitting data from surveys about sexual behaviour, but that can still be studied analytically, using mean field techniques. This allows us to obtain some general results about epidemic spreading. We show that using accumulated network data to estimate the static epidemic threshold leads to a significant underestimation of it. We also show that, for a dynamic network, the relative epidemic threshold is an increasing function of the infectious period of the disease, implying that the static value is a lower bound to the real threshold. 
\end{abstract}

\section{Introduction}
\label{Intro}
Even though the aim of mathematical modelling in epidemiology has always been to help predicting the patterns of spread of infectious diseases, the complexity of real populations has always constrained modellers to use strong assumptions. Even though these do not always guarantee the existence of analytic solutions, at least the models become {\em tractable}. On the other hand, the search for analitical simplicity, or beauty, has sometimes taken over more practical considerations.

One of the strongest assumptions used in most epidemiological models is the Law of mass action \cite{H}. First proposed by chemists, it postulates that in dynamical equilibrium the rate of a chemical reaction is proportional to the concentrations of the reactants, and can be derived from the probability of collision between reacting molecules. The analogy between the movements of molecules and living beings, drawn almost a century ago \cite{H}, leads to the epidemiological version of this postulate: the `force of infection' is proportional to the densities of infected and uninfected individuals (called `susceptibles' in the epidemiological literature). It implies assuming that the population has no structure, i.e. that every person can be in contact with every other (`random mixing').

In general, however, members of a population interact only with a very small subset of it. Thus, one way to go beyond the random mixing assumption is to consider that the members of the population form a social network. Its definition depends strongly on the type of interaction necessary to transmit the disease whose spread is being modelled. The advantage of this over the random mixing approach is that models can be better adapted to specific populations. Needless to say, this implies having more data about the social structure, as well as new concepts and tools to analyse them. Fortunately, these are provided by Social Network Analysis, a field that has developed rapidly in recent years \cite{WF}. The mathematics are not as straightforward as in the analysis of mass-action models, but for some cases some interesting results can be obtained by using approximations (some of them derived from statistical physics). One example is the simple relationship that exists for a disease with infectivity $\lambda$ and an infectious period $\alpha^{-1}$, between the relative epidemic threshold $\tilde \lambda_c=\lambda_c/\alpha$, and the topological properties of the network \cite{AM,P-SVb}:

\begin{equation}
\tilde \lambda_c= \frac{\langle x \rangle}{\langle x^2 \rangle}
\label{statthres}
\end{equation}

\noindent where $\langle x \rangle$ is the mean of the degree distribution of the social network, and $\langle x^2 \rangle$ is its variance.

Network epidemiology seems particularly well suited for the analysis of the spread of sexually transmitted diseases, as the definition of the network in this case is more straightforward (although not free of problems, see \cite{K}). The large number of surveys of sexual behaviour carried out in the last three decades has provided an invaluable resource for modellers. Interestingly, one common feature of many sexual networks built from survey data is that their degree distribution has a very long tail: there exist a small number of individuals who have a very large number of sexual contacts. Mathematically, this means that, even though $\langle x \rangle$ is rather small (typically less than 3), $\langle x^2 \rangle$ can be very large. Applying Eq.~(\ref{statthres}) to such networks (what, as explained below, is not altogether correct) would lead to the conclusion that, for those populations, even diseases with very low infectivity can trigger an epidemic. It has even been argued that some sexual networks have power law degree distributions with infinite variance \cite{LEASA,SMGFNAJG}, which would imply a vanishing epidemic threshold, but there is some controversy about this \cite{JH}.

One aspect that is usually disregarded in the network approach is the dynamic nature of social interactions. It is reasonable to assume that this dynamics produces a steady-state, in which the distribution of contacts does not change, even though at all times individuals are free to end their existing relationships and create new ones. Eq.~(\ref{statthres}) is derived for a static network, and is sometimes used to estimate the epidemic threshold of populations whose structure is deduced from sexual behaviour surveys. Respondents to these surveys, however, are usually asked about number of partners over a certain time period, and the distribution thus obtained is often used as a proxy for the steady state, or {\em instantaneous} distribution. But it is difficult to ascertain how close distributions of accumulated contacts can be to the instantaneous distribution \cite{N}. It is often suggested that if the time period asked about in the survey is similar to the infectivity period of the disease analysed, epidemic thresholds can be calculated by using the proxy network (see for example \cite{GJ,LEASA,HHM}). But in general this argument remains at a qualitative level. In principle, it should be possible to see whether the dynamics affects the spread of the disease only by generating a steady state distribution or there are other effects independent of this.

Models that take into account the dynamic nature of social network usually consider that the formation and dissolution of links between individuals are stochastic processes \cite{W}. More recently, such models have also been used to understand the spread of infectious diseases \cite{K1,TAGGME, DH,A,KE,VM}. But, in general, the additional complication of dealing with network dynamics has led either to models that have analytical solutions but that are too simple to be applied in a realistic setting, or to models that rely exclusively on numerical simulations, from which it is difficult to draw general conclusions. The model of network dynamics presented in the next section is an attempt at overcoming these limitations. It can be tailored to give similar accumulated degree distributions to those obtained in real surveys, as shown in the third section, but it also allows us to obtain some very general analytical results for the influence of network dynamics on the propagation of infectious diseases, using mean field techniques.

\section{Model}
\label{model}

We consider a population of $N$ individuals epidemiologically identical. As in this case it has been shown that static models with individuals placed on a bipartite network give identical predictions to models where the population is not divided into two groups \cite{N}, we have assumed that partnerships can 
be established between any two individuals. Thus, even though our model applies strictly only to homosexual populations, its predictions should be  qualitatively correct for heterosexual populations with similar epidemiological variables for both sexes.

Partnerhips can be established and dissolved with a rate that depends on features of the two individuals. As the only dynamic attribute we consider is the number of partners, we first assume that rates depend only on it. Thus, the rate of partnership creation between individuals $i$ and $j$ is $\rho(k_i,k_j,t)$ and the rate of partnership dissolution is $\sigma(k_i,k_j,t)$, where $k_i$ and $k_j$ are the number of current sexual partners of $i$ and $j$ at time $t$. As we only deal with steady states, hereafter the $t$ dependence is dropped from all quantities. 

In equilibrium, the master equation for the steady state degree distribution $P(k)$ becomes: 

\begin{eqnarray}
0 &=& -(N-1-k) P(k-1) \rho_{k-1} -  (k+1) P(k+1) \sigma_{k+1} \nonumber \\ & &
+ P(k) (N-1-k) \rho_k + k P(k) \sigma_k   
\label{equil}
\end{eqnarray}

\noindent where $\rho_k=\langle \rho(k,k_l) \rangle_l$ is the average probability that an individual with $k$ partners gets a new partner and $\sigma _k=\langle \sigma(k,k_l) \rangle_l$ is the average probability that an individual breaks one of his existing relationships. In principle, the link creation probability should be averaged only over those individuals that are not current partners of the individual. However, as in real populations $k$ is much smaller than $N$, this quantity is very well approximated by the average over the entire population:

\begin{equation}
\langle \rho(k,k_l) \rangle_l = N^{-1}\sum_{l=1}^N \rho(k,k_l) = \sum_{k_l}^N \rho(k,k_l) P(k_l)
\label{constrho}
\end{equation}

For the link dissolution probability, the distribution that should be used to calculate the average is $P(k_l|k)$, the degree distribution of the individuals that are connected to an individual having $k$ partners. However, if we assume that the dynamics does not generate a significant assortative mixing by degree, $P(k_l|k)$ can be written as $P(k_l|k)=k_l \, P(k_l)/ \langle k \rangle$. This is not a too stringent assumption, since there seems to be no definite tendency in mixing with respect to sexual activity: some sexual networks have been found to be weakly assortative \cite{GHASAWHH}, some neutral \cite{BMS} and some disassortative \cite{HGA}. The resulting average link dissolution is, then,

\begin{equation}
\langle \sigma(k,k_l) \rangle_l = \sum_l \sigma(k,k_l) k_l \, P(k_l)/ \langle k \rangle.
\label{constsig}
\end{equation}

Solving Eq.~(\ref{equil}) gives the steady state degree distribution:

\begin{equation}
P(k)=\frac{P(0)}{N^k} {N-1 \choose k} \prod_{i=0}^{k-1} \frac{\rho_i}{\sigma_{i+1}}
\label{steady}
\end{equation}

\noindent for $k>0$. $P(0)$ is obtained by normalizing the distribution. $P(k)$ can also be written as 

\begin{equation}
P(k;x_0, \cdots ,x_N)=\frac{P(0)}{N^k} {N-1 \choose k} \prod_{i=0}^{k-1} x_i
\label{steady2}
\end{equation}

\noindent where the $N$ parameters $x_i$ ($i=0, \cdots ,N-1$) are obtained by solving the $N$ self-consistency equations

\begin{equation}
x_i=\frac{\langle k \rangle \sum_l \rho(i,l) P(l;x_0, \cdots ,x_N)}{\sum_l \sigma(i,l) l \, P(l;x_0, \cdots ,x_N)}
\label{selfconsi}
\end{equation}

If a model is to be used for understanding the spread of a disease in a real population, its parameters should be adjusted by comparing with the available population data. For simpler models, it has been suggested that this could be done by using an empirical instantaneous distribution \cite{A}. In our model, however Eqs.~(\ref{steady}) and~(\ref{selfconsi}) show that rescaling the link creation and dissolution functions does not change the equilibrium distribution. This was to be expected, because changing the time scale cannot change the nature of the steady state reached. Thus, time scales should be obtained from other population measurements. An important problem of this approach is that, unfortunately, information about instantaneous degree distributions is usually {\em not} available. Instead, almost all surveys ask respondents about the number of sexual contacts accumulated over a certain time period. Thus, what we need to know from the model is the distribution of accumulated contacts (i.e. the probability of having had $k$ contacts during a given time period), $P_T (k)$, which can be written as

\begin{equation}
P_T (k)= \sum_{k'=0}^k P_T(k-k'|k')
\end{equation}

\noindent where $P_T(k-k',k')$ is the probability of having $k-k'$ {\em new} contacts over a time period of length $T$, conditional on having $k'$ partners at the beginning of that period. The equations that these conditional probabilities satisfy are

\begin{eqnarray}
\dot P_T(m|n) &=& \rho_n \, [P_T(m-1|n+1)-P_T(m|n)] + \nonumber \\
 & & n \sigma_n \, [ P_T(m|n-1)-P_T(m|n)]
\label{cumu}
\end{eqnarray}

\noindent for $0\leq m,n \leq N-1$, with $\rho_{N-1} \equiv 0$ and $\sigma_0 \equiv 0$. With the aid of some mathematical software, such as Mathematica or Matlab, this recursion can be solved exactly, for any desired value of $T$ (see Appendix). Using this, the parameters $\rho$ and $\sigma$ can be adjusted to fit the distributions obtained in any given survey. An example of this is given in the next section.

\section{Application examples}

The number of self consistency equations to be solved (Eqs.~\ref{selfconsi}) imposes a practical constraint on the models that can be effectively analized. One of the simplest ways to reduce the number of equations to only one is to consider functions of the form $\rho(k_i,k_j)=\rho(k_i) \rho(k_j)$ and $\sigma(k_i,k_j)=\sigma(k_i) \sigma(k_j)$. This choice has the added advantage of ensuring that there is no assortative mixing by degree. Note that if $\rho(k)$ is an increasing function of $k$, individuals with many partners are more likely to attract new ones. This is usually known as preferential attachment in the network literature \cite{BA}. Interestingly it has been shown that this is likely to play a role in the formation of sexual networks \cite{FdBSL}.

First we analyze two different models, called A and B, that generate almost the same instantaneous degree distribution. Model A is defined by the functions $\rho(k)=C_A k^3/(k+1)^2$ (for $k>0$), $\rho(0)=1$, and $\sigma(k)=1$, whereas model B is defined by $\rho(k)=C_B k^3/(k+1)$ (for $k>0$), $\rho(0)=1$, and $\sigma(k)=k$. $C_A$ and $C_B$ are numerical constants. The instantaneous distribution is $P(k)=P_0 D_k x^k/k^3$, where $D_k=\prod_{i=1}^k (1-i/N)$. $x$ is obtained by solving the self consistency equation for each model. The constants $C_A$ and $C_B$ are adjusted to obtain a degree distribution that has a mean value of order $1$, and a variance large enough to mimic the long tails observed in sexual networks. We find that there is a critical value for $C_A$ and $C_B$ below which the network is sparsely connected, and above which the network becomes dense, in the sense that each individual is connected to a significant fraction of the population (see Appendix). This is usually called a phase transition. Thus, to obtain a relatively wide degree distribution but keeping the network sparse, $C_A$ and $C_B$ were given values that are close to (but below) the critical value.

Fig. 1 shows that the mean field approach is a very good approximation for the corresponding stochastic model, both for the instantaneous degree distribution as well as for the accumulated ones. It also shows that, even for models with the same instantaneous degree distribution, the distribution of the number of  accumulated partners can be rather different. As a consequence, the usual approach of fitting the tail of these distributions with a power law function would not give the same exponent for models A and B. The accumulated distributions can be used to calculate epidemic thresholds, using Eq.~(\ref{statthres}), which can be considered as approximations to $\tilde \lambda^0_c$, the static threshold. The inset shows that these approximations can be very different from the actual value of $\tilde \lambda^0_c$.

\begin{figure} [!h]
\centerline{\includegraphics[width=.8\textwidth,clip=true]{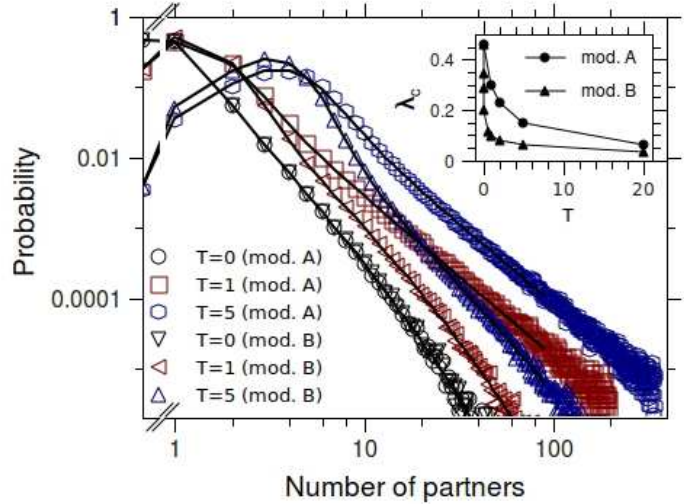}}
\caption{Distributions of number of partners accumulated over a time period $T$, for models A and B (see text). The full lines for $T=0$ are given by Eq.~(\ref{steady}), whereas the other lines are obtained by solving recursively Eqs.~(\ref{cumu}). Symbols correspond to simulations for a system with $10000$ individuals (averaged over 100 runs). The symbols and lines falling on the left vertical axis represent the fraction of individuals having $0$ sexual partners. Error bars are smaller then the symbols. The inset shows the static epidemic threshold calculated for the distribution of accumulated partners for different time periods, for both models.
} \label{figure1}
\end{figure}

To see whether these differences are relevant in a real setting, we have applied this model to data from the National Survey of Sexual Attitudes and Lifestyles II (NATSAL 2000), carried out in Britain in 2000-2001 \cite{NATSAL,JMECMW}. Participants were asked about the number of male and female partners during several, overlapping, time periods previous to the survey: 1 month, 3 months, 1 year, and 5 years. From these data, one can build, for each time period, the distribution of the number of  accumulated partners. 

Furthermore, we have only used the data related to homosexual men, since our model deals strictly with one-sex populations. However, as sexual orientation was not asked about to the participants of NATSAL, we have used a definition of MSM (men who have sex with men) as those men having reported at least one male partner within the five years prior to interview \cite{MFCWEMNMJ}. This leaves 166 out of 4762 male respondents. Because of recall problems, the accuracy of the reports decreases as the time period asked about increases \cite{CGMCG}. This is already apparent in the data for 5 years (not shown), where there is substantial heaping. In our case, this data set is further skewed because it has been used to define MSM. Thus, we have adjusted our model to fit only the degree distributions for 1 month, 3 months and 1 year (see Appendix). We have not used the data about lifetime number of partners, because the time periods involved were not the same for all participants (whose ages ranged from 16 to 44 years), as assumed in our model.

Fig. 2 shows the distribution of accumulated partners for the four time intervals analyzed. The fit is reasonably good for the three curves used. Even though the data for the 5 years period are overestimated, the tendency seems to be correct. The inset shows the approximations to the static threshold, calculated using the model degree distributions for several time periods (see Appendix). As in the previous figure, the approximations get worse when calculated using longer time periods. In fact, already the 1 month distribution leads to an underestimation of $\tilde \lambda_c^0$ of about 50 \%. 

To understand whether this underestimation is relevant, the spread of a disease should be analyzed taking into account the intrinsic dynamics of the network. The question is not only how close the real and static thresholds are, but even which one is larger, because it could happen that the real threshold was smaller than the static one, thus compensating for the underestimation of the approximations calculated with accumulated degree distributions. In the next section it is shown that this is not the case: real thresholds are always larger than static ones.

\begin{figure} %[!h]
\centerline{\includegraphics[width=.8\textwidth,clip=true]{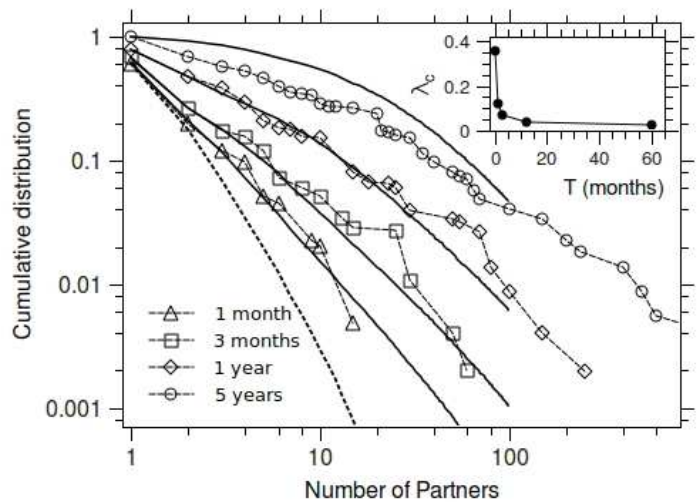}}
\caption{Cumulative distribution of the number of sexual partners accumulated over different periods of time for a population of homosexual men. Symbols correspond to data from the british National Survey of Sexual Attitudes and Lifestyles (NATSAL 2000). The lines joining the symbols are only guides to the eye. The full lines correspond to the fits of the epidemic model for each time period. The lowest dotted line is the prediction for the instantaneous cumulative degree distribution. The inset shows the values given by the model for the static epidemic threshold, calculated from the degree distribution for the time periods analyzed.
} \label{figure2}
\end{figure}

\section{Epidemic spread}
\label{epidemic}

We consider the propagation of a disease that can be cured, and that confers no immunity, i.e. individuals can be reinfected as soon as they become susceptible again. This type of models, called SIS, are considered acceptable models of sexually transmitted diseases as gonorrhea and chlamydia \cite{HY}.

It is assumed that, in an existing relationship between a susceptible and an infected individual, infection can pass with a probability $\lambda$ per unit time, and that infected individuals heal at a rate $\alpha$. We also assume that the social dynamics is not affected by the propagation of the disease. 
We need to calculate $P_x(k,I;t)$, the probability that at time $t$ an agent $x$ has $k$ simultaneous relationships and is infected. The master equation for this depends on the two point probabilities $P_{xy}(k_x,S; k_y,I; t)$, which in turn depend on three-point probabilities, and so on. To get a closed system we choose the simplest ansatz: $P_{xy}(k_x,S;k_y,I;t) \approx P_x(k_x \, S;t) P_y(k_y \, I;t)$. Using this, and averaging over all agents with the same number of partners, k, the master equation for $P(k,I)$ becomes

\begin{equation}
(\alpha \mathbf I + \mathbf A_\theta) \vec{P_I} = \overrightarrow {k P } \theta \, \lambda
\label{infsinmat}
\end{equation}

\noindent where $\mathbf A$ is a tridiagonal matrix defined by $(A_\theta)_{i \, i+1}=-i \sigma_i$, $(A_\theta)_{i \, i}=(N-i) \rho_{i-1} + (i-1) \sigma_{i-1} + (i-1) \lambda \, \theta$ and  $(A_\theta)_{i+1 \, i}=-(N-i) \rho_{i-1}$ and the vectors $\vec{P_I}$ and $\overrightarrow{k P}$ are given by $(P_I)_j=P(j,I)$ and $(kP)_j=j \, P(j)$. $P(j)$ is given by Eq.~(\ref{steady}). $\theta$ is the probability of having an infected partner \cite{P-SVb}, $\theta= \vec{k} \vec{P_I}/\langle k \rangle$, and is obtained from the self consistency condition,
 
\begin{equation}
\vec k (\alpha \mathbf I + \mathbf A_{\theta})^{-1} \overrightarrow {k P} = \frac{\langle k \rangle}{\lambda}.
\label{selfcon}
\end{equation}

The epidemic threshold can now be easily obtained by taking the limit $\theta \to 0$:

\begin{equation}
\lambda_c= \frac{\langle k \rangle}{\vec k (\alpha \mathbf I + \mathbf A_0)^{-1} \overrightarrow {k P}} 
\label{dynepithr}
\end{equation}

The fraction of infected individuals is 

\begin{equation}
n_I=\theta \, \lambda \, {\vec 1} (\alpha \mathbf I + \mathbf A_{\theta})^{-1} \overrightarrow {k P}
\end{equation}

\noindent where $\vec{1}$ is the vector with all components set to $1$. In the limit where the characteristic times of the disease are much shorter than the ones characterizing the social dynamics (i.e. $\lambda \to \infty$, $\alpha \to \infty$, but keeping $\tilde \lambda=\lambda/ \alpha$ constant), the usual result for a static network is obtained (Eq. 1): $\tilde \lambda_c^0= \langle k \rangle / \langle k^2 \rangle$. Intuitively one can think that the disease spreads so fast that it `sees' only the instantaneous network. The opposite limit can also be calculated (see Si text), giving $\tilde \lambda^{\infty}_c=\langle k \rangle^{-1}$. Thus in this case, the social dynamics is so fast that, in terms of disease spread, the network is equivalent to an `average' network where all nodes have the same degree, $\langle k \rangle$. Note that $\tilde \lambda^{\infty}_c >\tilde \lambda^{0}_c$. It is interesting to note that the social dynamics influences disease spread only through the instantaneous network of contacts, in the limit cases. 

Fig. 3 shows that the relative epidemic threshold of the NATSAL model is larger for diseases with larger infectious periods, $t_I=\alpha^{-1}$. Note that for infectious periods of the order of a few months, as is the case of untreated gonorrhea, chlamydia and syphilis, the difference between the corresponding threshold and the static approximation, $\tilde \lambda_c^0 $, can be significant. In terms of the nonnormalized epidemic threshold, the inset of Fig. 3 shows that when the dynamics of the network is taken into account, $\lambda_c$ decreases more slowly with $t_I$.

\begin{figure}%[h]
\centerline{\includegraphics[width=.8\textwidth,clip=true]{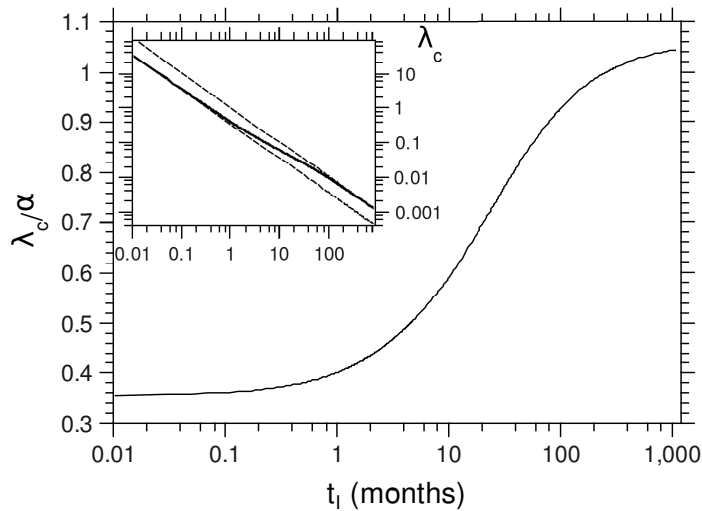}}
\caption{Relative epidemic threshold as a function of the infectious period of the disease. The inset shows the epidemic threshold as a function of infectious period. Both curves were calculated using the model obtained by fitting the NATSAL data (NATSAL model).} 
\label{figure3}
\end{figure}

Interestingly, it can be proved (see Appendix) that the effect of the dynamics is the same for {\em all} possible choices of the link creation and dissolution functions, $\rho(k_i,k_j)$ and $\sigma(k_i,k_j)$: the relative epidemic threshold always grows monotonously with $t_I$. Even though the mean field approximation is not very good for sparse networks (as should be the case of most instantaneous sexual networks), it can be conjectured that the picture is not qualitatively different. This is supported by simulations carried out for the stochastic analog of the NATSAL model. Fig. 4 shows that the qualitative behavior of the simulation curves is well predicted by the mean field approximation. Note that the real epidemic threshold is even larger than the mean field value and therefore the underestimation mentioned before is even worse when compared with simulation values.

For large values of the infectivity, Fig. 4 shows that $n_I$, the fraction of infected individuals in the endemic state, grows with $t_I$. This too is a general feature of this kind of models. Interestingly, for very large $\lambda$, $n_I$ does not tend to $1$:

\begin{equation}
\lim_{\lambda \to \infty} n_I = 1- \frac{P(0)}{1+t_I \langle \rho_0 \rangle} 
\end{equation}

\noindent In a static network (i.e. $t_I \to 0$), the disease cannot reach isolated individuals. In the dynamic case, however, even momentarily isolated people get a partner after a time $1/\langle \rho_0 \rangle$, on average. But there is a probability that isolated, infected people get cured before they get a partner. This ensures that there is always a fraction of the isolated individuals that is not infected, no matter how high the infectiousness of the disease is. The proportion of partners that are infected, $\theta$, is also an increasing function of $\tilde{\lambda}$ but it tends to $1$ for large infectivities, for all values of $t_I$.  It can also be proved that, for fixed values of $\tilde{\lambda}$, $\theta$ is a decreasing function of $t_I$.

\begin{figure}%[h]
\centerline{\includegraphics[width=.8\textwidth,clip=true]{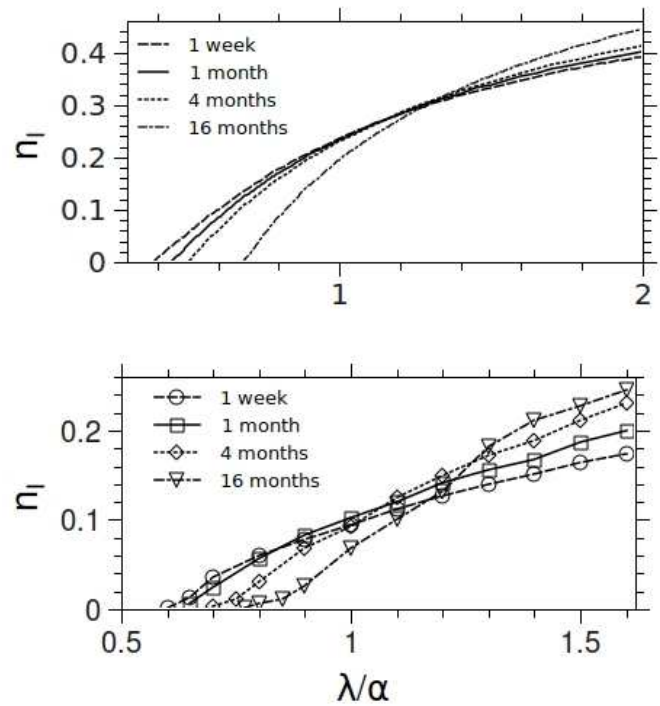}}
\caption{Fraction of the population that is infected, in the equilibrium state, as a function of the infectivity of the disease, for several values of the infectious period. The upper panel shows the theoretical curves obtained for the NATSAL model. The lower panel shows the results of simulations of populations of $10000$ individuals, using the same paramters as for the NATSAL model. Symbols correspond to averages over $100$ runs. The lines joining the sysmbols are only guides to the eye.} \label{figure4}
\end{figure}

\section{Including intrinsic features and neighbourhoods}
\label{intrinsic}

The model analyzed in the previous sections can be extended in many ways, in order to make it more realistic. One of them is to consider that the attraction between individuals can depend not only on the number of partners, which is a dynamical variable, but also on intrinsic features of each individual, called {\em fitness} in the network literature, that do not change over time (or at least over the times relevant for the problem). Many characteristics have been proposed to account for attraction, as beauty, talent, socioeconomical status, and even geographical location. The downside to this added realism is that such features are not easy to univocally define \cite{Ho}, let alone quantify. It is interesting, however, to see that some general properties can be derived for our model.

We assume that the fitness $f$ takes a finite number of values, whose probability mass function is $\Pi(f)$. The rates of partnership creation and dissolution depend now on the $f$ of each agent: $\rho(k_i,f_i,k_j,f_j)$ and $\rho(k_i,f_i,k_j,f_j)$. The population can be divided in subpopulations with a common value of $f$, with a degree distribution $P(k|f)$ given by Eq.~(\ref{steady}). One important difference with the model analyzed in the previous sections is that the time average of the number of partners is not the same for all individuals, but depends on their fitness. The interaction between the subpopulations is encoded in the self consistency parameters $x_i(f)$, calculated from

\begin{equation}
x_i(f)=\frac{\langle k \rangle \sum_l \rho(i,f;l,f_2) P(l;x_0, \cdots ,x_N|f_2)}{\sum_l \sigma(i,f;l,f_2) l \, P(l;x_0, \cdots ,x_N|f_2)}
\label{selfconsi2}
\end{equation}

It is also possible to obtain the distribution of accumulated contacts. In this case $P_T(m|n,f)$ is the probability that an individual with fitness $f$, having $n$ partners at the beginning of a given time period of duration $T$, has had $m$ partners at the end of that period. There is now a set of equations for each $f$, analog to Eqs.~(\ref{cumu}), that can be solved independently of each other. The degree distribution for the period $T$ is $P_T(m)=\sum_{f} \sum_{n=0}^m P_T(m|n,f)P(n)\Pi(f)$. 

The analysis of the spread of an infectious disease can be carried out much in the same way as in the previous section. The mean field approach leads to an equation analog to Eq.~(\ref{infsinmat}), for each subpopulation. The probability that a partner of an individual is infected, $\theta$, is again assumed to be independent of the individual, and is obtained by solving:

\begin{equation}
\langle \vec k (\alpha \mathbf I + \mathbf A_{\theta})^{-1} \overrightarrow {k P_f} \rangle_f = \frac{\langle k \rangle}{\lambda}
\label{selfcon2}
\end{equation}

\noindent where $(k P_f)_j=j \, P(j|f)$, $\langle \rangle_f$ denote an average over the distribution $\Pi(f)$ and $\langle \rangle$ denotes an average over both $\Pi(f)$ and $P(n)$. The epidemic threshold in this approximation is 

\begin{equation}
\lambda_c= \frac{\langle k \rangle}{\langle \vec k (\alpha \mathbf I + \mathbf A_0)^{-1} \overrightarrow {k P} \rangle_f} 
\label{thresh2}
\end{equation}

It is instructive to compare the cases where different fitness distributions generate the same instantaneous network. As expected, the static limit ($t_i \to 0$) does not depend on $\pi(f)$. But the opposite limit does depend on the fitness:

\begin{equation}
\tilde \lambda_c^{\infty} (\Pi(f))= \frac{\langle k \rangle}{\langle \left( \overline{k(f)} \right)^2 \rangle_f} 
\label{threshlim}
\end{equation}

\noindent where $\overline{k(f)}$ is the average of $k$ over the individuals with the same value of $f$. If there is a nontrivial fitness distribution, it can be shown that this value is strictly smaller than $\tilde \lambda_c^{\infty}=1/\langle k \rangle$, the limit found in the previous section. In other words, the effect of the social dynamics on the spread of the disease is less pronounced if the instantaneous network is (at least partly) generated by the features of the individuals. 

In STD epidemiology it is often assumed that there is a small group of individuals, usually called {\em core group}, whose contribution to the spread of the disease is disproportionately large. Even though there is some ambiguity in the exact characterization of it \cite{TT}, this label is frequently applied to people with very many sexual contacts \cite{A}. Our result suggests that, even having the same number of individuals at any time, dynamic core groups (whose composition changes with time) might be not as effective as static ones in driving an epidemic.

One potential drawback of including intrinsic features is that the computational work needed to obtain the different predictions of the model is multiplied by the number of possible values of the fitness. It must be noted, however, that in sociological studies many features are quantified with a very small number of values. For example, income is usually quantified in quintiles or deciles, and physical attractiveness, because of its intrinsic ambiguity, has been quantified in many sociological studies in scales having between 5 and 10 values.

Another aspect of the model that can be criticized is that, at any given time, any two individuals in the population can become sexual partners. This is not only geographically but also (and even more) socially not realistic. One way to overcome this limitation is to assume that each individual can only become a sexual partner of a fixed set of individuals, which form his or her `social neighbourhood'. Numerical simulations show that, for populations with neighbourhoods consisting of a few hundred individuals, results are almost indistinguishable from the ones presented in the previous sections. 

\section{Discussion and conclusion}
\label{conclusions}

Most models that take social dynamics into account seem to belong to two groups. One group consists of models that are 
analytically solvable but are too schematic to account for many important features of real populations. The other group consists of models that are much more complex, with many parameters that can be obtained from population data, but whose very complexity implies that their study can only be carried out by means of computational simulations. The model presented here is an attempt at bridging the gap between these two groups. On the one hand, it is sufficiently general to allow its parameters to be obtained by fitting data from population surveys. The example analyzed shows that the fits obtained can be very reasonable. On the other hand, the model can be studied analytically  using mean field techniques, which allows us to obtain some general results.

We have found that, because of the interplay between the social and the epidemic dynamics, the relative epidemic threshold, as a function of the average duration of infection, increases monotonically between the two limit cases, $\tilde \lambda_c^0=\langle k \rangle / \langle k^2 \rangle $ and $\tilde \lambda_c^{\infty}=1 / \langle k \rangle $. Thus, approximating the epidemic threshold by the static network threshold, entails an underestimation. And the example analized shows that, in real cases, this underestimation can be significant for diseases having an infectious period of the order of months. But, even in the case when $\tilde \lambda_c^0$ is a good approximation, the problem that remains is how to estimate its value from survey data. Participants in surveys about sexual behaviour are usually asked about number of partners during one or several time periods. Any properties of the instantaneous contact network must therefore be inferred from that information. Usually, $\tilde \lambda_c^0$ is estimated from the network built by considering the distribution of the number of accumulated partners as a degree distribution, for each time period. We have shown that, as is usually assumed, this approximation improves as shorter time periods are considered. Unfortunately, we have also shown that, in real cases, even the values obtained for rather short time periods  (1 month) can be much smaller than $\tilde \lambda_c^0$.

It is often assumed that to study the spread of diseases with short infectious periods the relevant information is encoded in the distribution of sexual partners for small time periods, whereas longer time periods (of the order of years) are more relevant for diseases with long infectious periods. The results of the previous sections show that this might not be the case, at least for the epidemic threshold. It is true that sometimes this threshold is well approximated by the static limit, whose estimation necessitates information about sexual partners in time periods as short as possible. But for diseases with long infectious periods, we find that the epidemic threshold obtained with distribution of partners for long time periods underestimates the static epidemic threshold, which in turn underestimates the real value. Therefore, for this kind of diseases, the best would be to to build a good social dynamics model by fitting the empirical data for several time periods, and to calculate its corresponding epidemic threshold.

Dynamic models as the one presented here still need the addition of many features before being considered as reasonable representations of real populations, such as the possibility of having asymptomatic individuals, and the division of the population into groups with different epidemiological characteristics. There is also room for improvement in the approximations used for the analysis of the model. One possibility is to go one step further from the mean field theory and to consider a pair approximation. It is not clear, however, whether such modifications would lead to a model amenable to analytical solutions or approximations, which is one of the main advantages of the model presented in this paper. 

\section{Acknowledgments}
I wish to thank M.N. Kuperman and D. H. Zanette for a critical reading of the manuscript and useful suggestions.

\vspace{1cm.}

\appendix

\noindent {\LARGE \bf Appendix}

\section{Distribution of accumulated contacts}
\label{appendixa}

By Laplace transforming Eqs. (9), solving, and back transforming, it can be shown that the probabilities that an individual has had $m$ new contacts at the end of a time period of length $T$, given that he had $n$ at the beginning of that period, are of the form:

\begin{equation}
P_T (m|n) = \sum_{i=0}^m \sum_{j=0}^{m+n-i} A_{ij}^{mn} \frac{T^i \mathrm{e}^{-T(\rho_j+j \sigma_j)}}{i!} 
\end{equation}

The constants $ A_{ij}^{mn}$ are obtained from the following recursions:

\begin{eqnarray}
A_{ij}^{mn}(c_n-c_j)+ A_{i+1 \, j}^{mn} &=& \rho_n  A_{ij}^{m-1 \, n+1} + n \sigma_n  A_{ij}^{m \, n-1} \nonumber \\ & & \mbox{for $i=1, \cdots, m-1$ and $j=0, \cdots ,m+n-i-1$}\nonumber \\[10pt]
A_{i \, m+n-i}^{mn}(c_n-c_{m+n-i}) &=& \rho_n A_{i \, m+n-i}^{m-1 \, n+1} \,\,\,\mbox{for $i=1, \cdots, m-1$} \nonumber \\[10pt]
A_{m j}^{mn}(c_n-c_j) &=& n \sigma_n A_{m \, j}^{m \, n-1} \,\,\,\,\mbox{for $j=0, \cdots,n-1$}
\end{eqnarray}

The remaining constants are obtained from the conditions $P_0(m|n)=0$ if $m>0$ and $P_0(0|n)=1$: $A_{0m}^{mn}=-\sum_{j \neq m}^{m+n} A_{0j}^{mn}$, $A_{00}^{0n}=1-\sum_{j=0}^{n-1} A_{0j}^{0n}$, and $A_{00}^{00}=1$.

\section{Models A and B}
\label{appendixb}

For models of the form $\rho(k_i,k_j)=\rho(k_i) \rho(k_j)$ and $\sigma(k_i,k_j)=\sigma(k_i) \sigma(k_j)$ the self consistency parameters are $x_i=\frac {\rho_i}{\sigma_{i-1}} x$. $x$ is obtained by solving

\begin{equation}
x=\langle k \rangle \frac{\sum_l \rho(i,l) P(l;x)}{\sum_l \sigma(i,l) l \, P(l;x)}.
\label{selfconsAp}
\end{equation}

If now all the creation functions are multiplied by the same constant, $C$, and the self consistency parameter is rescaled as $x'=Ax$, Eq. \ref{selfconsAp} becomes 

\begin{equation}
\frac{x'}{C}= \frac{\langle k \rangle \sum_l \rho(i,l) P(l;x')}{\sum_l \sigma(i,l) l \, P(l;x')} \equiv f(x')
\label{selfconsAp2}
\end{equation}

\begin{figure} [!h]
\centerline{\includegraphics[width=10cm,clip=true]{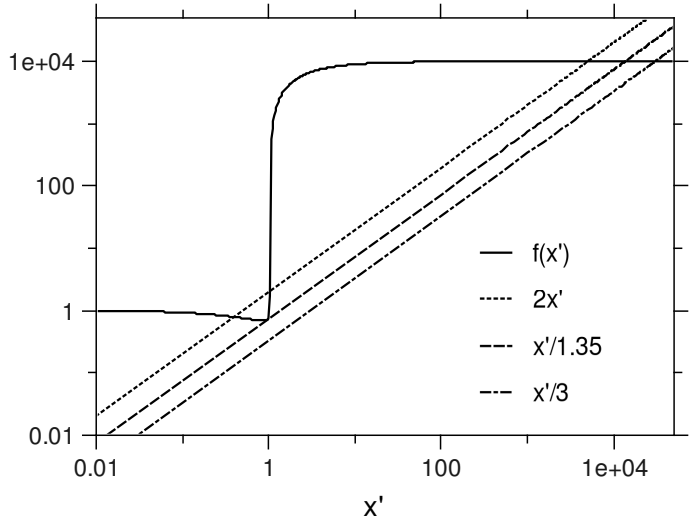}}
\caption{$f(x')$ as a function of $x'$ for model $A$ (see text).} 
\label{figureAP}
\end{figure}

As mentioned in the text, models A and B are defined as follows. Model A: $\rho(k)=C_A k^3/(k+1)^2$ (for $k>0$), $\rho(0)=1$, and $\sigma(k)=1$. Model B: $\rho(k)=C_B k^3/(k+1)$ (for $k>0$), $\rho(0)=1$, and $\sigma(k)=k$. $C_A$ and $C_B$ are numerical constants. The instantaneous distribution is $P(k)=P_0 D_k x^k/k^3$, where $D_k=\prod_{i=1}^k (1-i/N)$. Fig.\ref{figureAP} shows $f(x')$ and $x'/C$, for different values of the constant $C$, for $N=10000$. At $A \approx 1.35$ there is a discontinuous phase transition from a network with $x' \approx 1$ to a network with $x'=O(N)$. 

\section{NATSAL model}
\label{appendixc}

To obtain a model that fits the NATSAL data we have taken into account the fact that the number of respondents was rather small and, as a consequence, the sampling error for the number of repondents declaring having had more than two partners is likely to be rather large. We have chosen to adjust separately only the values of $\rho(0)$, $\rho(1)$, $\sigma(1)$, and $\sigma(2)$ to fit the number of respondents that reported 0 or 1 partner. The rest of the data were fitted using the generic functions $\rho(i)=C_{\rho} i^{\beta}/(i+1)^{\beta-2}$ and $\sigma(i)=C_{\sigma} i^{\gamma}$. The fits were performed sequentially. In the first step we fitted $\rho(0)$, $\rho(1)$, $\sigma(1)$, and $\sigma(2)$ using the analytic expressions for $P_T(0)$ and $P_T(1)$. In the second step, a coarse sampling of parameter space was performed, in order to select a suitable region on which to focus. This selection was performed by calculating several different distributions $P_T(k)$ for relatively small values of $k$ ($\approx 50$) (which takes only a few seconds of computation time) and choosing the one that best fitted the data. In the third step, a fine tuning of the parameters found was performed by generating some `full' distributions (up to $k=100$) (which takes tipically a couple of days of computation time) for small displacements from the parameters selected in the previous step. The values obtained for the different parameters are given in Table \ref{tableNATSAL}.

To compensate for the heaping present in the number of partners reported (i.e. the preference of respondents for round numbers, specially for large numbers), we have applied geometric binning to the data. Nevertheless, the fits obtained are quite good for other presentations of the data, as the cumulative numbers of partners (see Fig. 2 in the main article)

The estimates of the static epidemic threshold shown in the inset of Figs. 1 and 2 in the main text were calculated using the accumulated partners distributions found, i.e. up to $k=100$. Therefore the values are not exact, but it can be shown that they are upper bounds to the values calculated using the full distributions. This means that the difference between the exact estimations and the static threshold is even larger than what is shown in the insets.

\begin{table}[ht!]
\centering
\caption{Parameters used for the NATSAL model}
\begin{tabular*}{\hsize}{@{\extracolsep{\fill}}lcr}
\hline
$\rho_0$&$0.0373$\cr
$\rho_1$&$0.0459$\cr
$\sigma_1$&$0.0348$\cr
$\sigma_2$&$0.0842$\cr
$\beta$&$2.7$\cr
$\gamma$&$1$\cr
$C_{\rho}$&$0.0229$\cr
$C_{\sigma}$&$0.132$\cr
\hline
\end{tabular*}
\label{tableNATSAL}
\end{table}

\section{Properties of the epidemic threshold}
\label{appendixd}

Using Eq. 5 of the main text, the elements of matrix $\mathbf A_0$ can be written as $(A_0)_{i \, i+1}=-i \sigma_i$, $(A_0)_{ii}=\frac{P(i)}{P(i-1)} i \sigma_i + (i-1) \sigma_{i-1}$, and $(A_0)_{i \, i-1}=- \frac{P(i-1)}{P(i-2)}(i-1) \sigma_{i-1}$. If we define a diagonal matrix $\mathbf D_P$ such that $(D_P)_{ii}=P(i)$, it is straightforward to see that $\mathbf A_0$ can be written as ${\mathbf A_0}= {\mathbf A'_0} \mathbf D_P^{-1}$, where $\mathbf A'_0$ is a symmetric, tridiagonal matrix, with vanishing row (and column sums), defined by $(A'_0)_{i \, i+1}=-i P(i) \sigma_i$. Therefore, Gershgorin theorem implies that $\mathbf A'_0$ is positive-definite. That is, it has the property that 

\begin{equation}
x^t \mathbf {A'_0} x \geq 0
\label{eq1appB}
\end{equation}

\noindent for any vector $x$. Using $\vec x=\mathbf{D_P}^{-1} (\alpha \mathbf{I}+ \mathbf{A_0})^{-1} \mathbf{D_P} \vec k$ in Eq. \ref{eq1appB} and using the definition of $\lambda_c$ (Eq. 12) it can be shown that 

\begin{equation}
\frac{\partial \lambda_c}{\partial t_I} = {\vec k} (\alpha \mathbf I + \mathbf A_0)^{-1} {\mathbf A_0} (\alpha \mathbf I + \mathbf A_0)^{-1} {\mathbf D_P} {\vec k} = x^t \mathbf {A_0 D_P} x \geq 0
\end{equation}

We can also show that the growth of $\lambda_c$ is not unbounded, by calculating $\lim_{t_i \to \infty} \tilde \lambda_c=\lim_{\alpha \to 0} \tilde \lambda_c$. For this, we need to calculate the limit of $(\mathbf{A_0}+\alpha \mathbf{I})^{-1}$. Note that it can be written as $(\mathbf{A_0}+\alpha \mathbf{I})^{-1}=\mathrm{adj} ((\mathbf{A_0}+\alpha \mathbf{I}))/  \det((\mathbf{A_0}+\alpha \mathbf{I}))$. The adjoint of a matrix $\mathbf{A}$ is defined as $(\mathrm{adj} \mathbf{A})_{ij} =(-1)^{i+j} M_{ij}$, where $M_{ij}$ are the minors of $\mathbf{A}$, i.e. $M_{ij}$ is the determinant of the matrix obtained by deleting row $i$ and column $j$ from $\mathbf{A}$.

The minors of $\mathbf{A_0}$ can be written as $M_{ij}=M'_{ij} P(j)/ \det(\mathbf{D_P})$, where $M'_{ij}$ are the minors of $\mathbf{A'_0}$. But the fact that all row and column sums vanish implies that $M'_{ij}=(-1)^{i+j} M'_{11}$. It also implies that the determinant of $\mathbf{A_0}+\alpha \mathbf{I}$ can be calculated by replacing each element of its first row by $\alpha$. Using the Laplace expansion for the determinant, we then get 

\begin{equation}
\det(\mathbf{A_0}+\alpha \mathbf{I})=\alpha \sum_{j} (-1)^{1+j} M_{1j} + O(\alpha^2) = \alpha \frac{M_{11}}{\det \mathbf{D_P}} + O(\alpha^2)
\end{equation}

Using now that $(\mathrm{adj} \mathbf{A_0})_{ij}= M_{11} P(j)/ \det \mathbf{D_P}$, we obtain

\begin{equation}
\lim_{\alpha \to 0} \alpha^{-1} ((\mathbf{A_0}+\alpha \mathbf{I})^{-1})_{ij}=P(j)
\end{equation}

Replacing now this expression in Eq. 12 leads to $\lambda^{\infty} = \lim_{\alpha \to 0} \tilde \lambda_c=1/\langle k \rangle$.

\end{document}